\documentclass[journal]{IEEEtran}
\usepackage{amsmath,amsfonts}
\usepackage{algorithmic}
\usepackage{algorithm}
\usepackage{array}
\usepackage[caption=false,font=normalsize,labelfont=sf,textfont=sf]{subfig}
\usepackage{textcomp}
\usepackage{stfloats}
\usepackage{url}
\usepackage{verbatim}
\usepackage{graphicx}
\usepackage{cite}
\begin{document}

\title{Surface Protection and Activation of Mid-IR Plasmonic Waveguides for Spectroscopy of Liquids}

\author{Mauro David, Ismael C. Doganlar, Daniele Nazzari, Elena Arigliani, Dominik Wacht, Masiar Sistani, \\ Hermann Detz, Georg Ramer, Bernhard Lendl, Walter M. Weber, Gottfried Strasser, and Borislav Hinkov

\thanks{M. David, I. C. Doganlar, D. Nazzari, E. Arigliani, M. Sistani, H. Detz, W. M. Weber, G. Strasser, and B. Hinkov are with the Institute of Solid State Electronics \& Center for Micro- and Nanostructures, TU Wien, Vienna, Austria (e-mail: mauro.david@tuwien.ac.at ; borislav.hinkov@tuwien.ac.at).} 
\thanks{D. Wacht, G. Ramer, and B. Lendl are with the Institute of Chemical Technologies and Analytics, TU Wien, Vienna, Austria.} 
\thanks{H. Detz is with CEITEC, Brno University of Technology, Brno, Czech Republic}}

\maketitle

\begin{abstract}
Liquid spectroscopy in the mid-infrared spectral range is a very powerful, yet premature technique for selective and sensitive molecule detection. Due to the lack of suitable concepts and materials for versatile miniaturized sensors, it is often still limited to bulky systems and offline analytics. Mid-infrared plasmonics is a promising field of current research for such compact and surface-sensitive structures, enabling new pathways for much-needed photonic integrated sensors. In this work, we focus on extending the concept of Ge/Au-based mid-infrared plasmonic waveguides to enable broadband liquid detection. Through the implementation of high-quality dielectric passivation layers deposited by atomic layer deposition (ALD), we cover the weak and water-soluble Ge native oxide. We show that approximately 10 nm of e.g. Al$_2$O$_3$ or ZrO$_2$ can already protect the plasmonic waveguides for up to 90 min of direct water exposure. This unlocks integrated sensing schemes for broadband molecule detection based on mid-infrared plasmonics. In a proof-of-concept experiment, we further demonstrate that the ZrO$_2$ coated waveguides can be activated by surface functionalization, allowing the selective measurement of diethyl ether at a wavelength of 9.38 $\mu$m. 
\end{abstract}


\section{Introduction}
\IEEEPARstart{T}{he} mid-infrared (mid-IR) spectral region is highly suitable for a wide variety of applications including optical free-space communication \cite{Dely2022,Hinkov2019a,Corrias2022} and spectroscopy of molecules in the gas \cite{Wysocki2010,Spagnolo2012}, liquid \cite{Alcaraz2017,Bibikova2017,Hinkov2022} and solid phase \cite{Fuchs2010}. Investigating liquids recently gathered particular interest, sparked by significant progress in the understanding and control of the spectral properties of quantum cascade (QC) devices \cite{Hinkov2008,Heydari2015,Suess2016,Dely2022} leading to novel high-performance QC lasers (QCLs) and detectors (QCDs) \cite{Bai2011b,Schwarz2017,Jollivet2018,Bigioli2020,Delga2020}. They include high-power QC lasers (QCLs), enabling the penetration of thicker sample films on the hundreds of micrometer scale \cite{Schwaighofer2016} and novel-wavelength and high-performance QC detectors (QCDs) \cite{Marschick2022}. This allows monitoring wide concentration ranges characterized by sensor linearity and high saturation thresholds \cite{Dabrowska2022}. But even those new and other comparable improved devices still require rather bulky experimental setups relying on external components including flow cells \cite{Schwaighofer2016,Lopez-Lorente2017,Dabrowska2022} and in some cases also need specific sample preparation techniques \cite{Schwaighofer2016,Bibikova2017,Dabrowska2022}. The next generation of mid-IR liquid sensors characterized by compact footprint and robust operation is still in its infancy. It demands for breakthrough integration concepts for realizing miniaturized photonic integrated circuit (PIC) devices. Previous investigations could demonstrate that novel mid-IR plasmonic approaches are suitable for fingertip-sized monolithic PICs \cite{Schwarz2014}. They can be further exploited in terms of lab-on-a-chip sensors allowing highly-sensitive and -selective liquid spectroscopy measurements \cite{Schwarz2014,Hinkov2022}, when merging same-wavelength operating QCLDs with suitable mid-IR plasmonic materials and waveguide geometries \cite{Schwarz2014}. \\
While, similar to regular QCLs, the concept of integrated QCLDs has no fundamental limitations concerning its wavelength coverage throughout the entire mid-infrared spectral range, suitable (broadband) plasmonic materials and concepts in this wavelength range are rather scarce. Therefore, finding alternative pathways to make existing plasmonic approaches and material systems suitable for a much wider range of applications is very compelling. Existing octave-spanning and low-loss mid-IR Ge/Au semiconductor-loaded surface plasmon polariton (SLSPP) waveguides \cite{David2021}, are highly suitable candidates for this task. While showing an excellent trade-off between mode confinement and propagation length, their topmost layer, which is Ge oxide, easily dissolves in water \cite{Xie2012,Berghuis2021} even at moderate temperatures, deteriorating any possible sensor performance. \\
In this work we show, that prominent "high-$k$ dielectrics" including ZrO$_2$ and Al$_2$O$_3$, are suitable for surface protection of Ge-SLSPP waveguide sensors towards direct water exposure. The protective layers are deposited by atomic layer deposition (ALD) with a thickness of a few nanometers only. To investigate their impact on the plasmonic properties, we show results from finite element method (FEM-)based simulations and compare them to optical measurements performed with and without ALD coating in a custom-made waveguide characterization setup. Final confirmation of the suitability of our coating approach is obtained by exposing coated and uncoated Ge/Au SLSPP waveguides to water for up to 90 minutes. Re-measuring the waveguide losses together with monitoring the changes in waveguide profile and analyzing the surface roughness by atomic force microscope (AFM) demonstrates the protective capabilities of the coatings. In an additional proof-of-concept experiment, we fully open the pathway towards on-chip liquid sensing, through surface activating the ZrO$_2$-coated waveguides with a trimethylsilyl functionalization. This allows us to selectively measure diethyl ether on the waveguide surface at a wavelength of 9.38 $\mu$m, followed by its evaporation resulting in a fully recovered plasmonic surface. The repeatability of the measurement is also confirmed.

\section{Plasmonic Concepts in the mid-IR Spectral Range}
Surface plasmon polaritons (SPPs) are the collective electron density oscillations, observed at an interface with a sign-change of the dielectric function, e.g. between a metal and a dielectric \cite{Sarid1981}. They combine the high-speed capabilities of photonic circuits with the ability of miniaturization below the diffraction limit for visible \cite{Guo2013,Huang2018,Sistani2019} to near-IR wavelengths \cite{Huang2018,Sistani2019}, provided through plasmonic confinement \cite{Barnes2003,Ozbay2006,Falk2009}. SPPs can be confined and guided along electrical structures including wires \cite{Falk2009,Guo2013} and waveguides \cite{Ozbay2006,Lal2007,Guo2013,David2021}. While this results in relatively high guiding losses \cite{Guo2013}, it still enables implementation in ultra-confined and high-speed SPP and localized SP (LSP) detectors \cite{Huang2018,Sistani2019,Sistani2020}. \\
In contrast, mid-IR plasmonics is still pretty much in its infancy. The traditional noble metal-based structures show poor mode confinement in the mid-IR, with modes extending far beyond the wavelength scale into the dielectric medium \cite{Law2013a,Schwarz2014,David2021}. This makes them highly unsuitable for plasmonic on-chip and mode-guiding applications. Alternatives, such as heavily doped epitaxial group IV and III-V semiconductors, transition metal nitrides, transparent conductive oxides, (metal-)silicides or graphene have been realized in recent years \cite{Zhu2008,Cleary2015,Zhong2015,Taliercio2019}. But they still show certain limitations. For example, highly doped epitaxial semiconductors benefit from tailoring the plasma frequency through adjusting their doping level, but they require high-quality epitaxial growth processes and are limited to suitable underlying substrate materials. A relatively simple way to overcome performance limitations in mid-IR plasmonics is to realize so-called dielectric-loaded surface plasmon polariton (DLSPP) waveguides \cite{Schwarz2014}. They enable mode confinement on the wavelength scale and guiding from 10's to 100's of micrometer along the chip surface, as e.g. shown for simple SiN/Au structures \cite{Schwarz2014,Hinkov2022}. When thin dielectric slabs ($\sim$200--300 nm) are applied, the plasmonic mode mostly ($>$96\%) propagates in the surrounding dielectric medium like e.g. air, making this approach highly suitable for, e.g. real-time \textit{in-situ} sensing applications in liquids, as previously demonstrated for proteins measured around 6.2~$\mu$m wavelength \cite{Hinkov2022}. Unfortunately, the covered wavelength range of SiN as dielectric loading material is limited due to absorptions between 7 - 16 $\mu$m \cite{Kischkat2012}.

\section{Semiconductor/Metal Structures for Plasmonics in the mid-IR Spectral Range}
\begin{figure}[ht]
						\centering
						\includegraphics[width=0.9\linewidth]{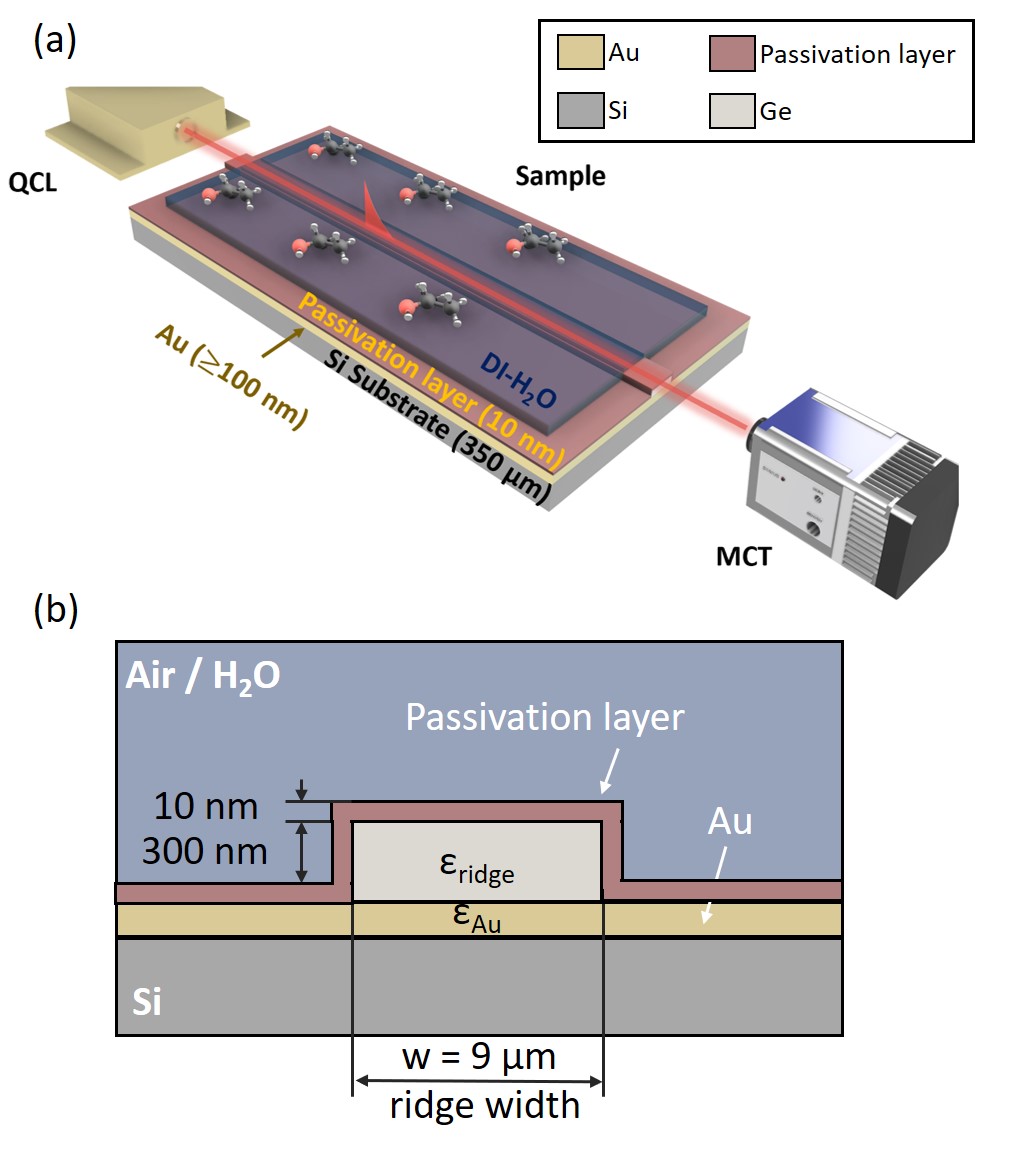}
						\caption{(a) Sketch of the sensing concept. The figure shows the plasmonic waveguide submerged in deionized (DI) H$_2$O together with an analyte, and measured with an externally coupled quantum cascade laser source and Mercury-Cadmium-Telluride (MCT) detector. The individual 
						layers including the one for surface passivation are indicated. (b) Cross-sectional profile of the whole waveguide layer sequence including the typical device dimensions.}
						\label{Waveguide+Sim}
\end{figure}

One possibility to extend the operational range of mid-IR DLSPP waveguides is by exchanging their dielectric-loading layer with another, better suitable material. For this, we recently realized a novel concept based on SLSPP waveguides. We use Ge \cite{David2021} as a highly transparent material in the entire mid-IR spectral range. In addition, it can be easily deposited in a suitable quality by a regular sputtering process and patterned by following state-of-the-art cleanroom fabrication techniques. This approach enables the realization of octave-spanning plasmonic waveguides from 5.6 - 11.2 $\mu$m \cite{David2021}. Its experimental demonstration is so far only limited to the used excitation laser source and can in principle be further extended to cover the whole range between 2 - 14 $\mu$m. \\
However, Ge is not very resilient against the exposure to liquids. Simultaneously, there is a lack of other more stable materials with similar optical properties as Ge, especially in the spectrally narrower long-wave infrared (LWIR, $\sim$8~-~12~$\mu$m) \cite{Lin2018}. Therefore, as alternative approach, we enhance the robustness of our Ge/Au plasmonic waveguides by applying appropriate ALD surface passivation coatings to extend the Ge-SLSPPs operational range to liquids. The group IV semiconductor Ge already gathered particular attention in the past, as an intermixed material with Silicon (Si) in traditional MOSFETs \cite{Goley2014,Wind2021}. It is fully compatible to complementary metal-oxide-semiconductor (CMOS) fabrication processes. Enhanced performance was enabled by the implementation of high-$k$ dielectrics, substituting SiO$_2$ as gate material to prevent limitations from shrinking transistor dimensions \cite{Xie2012}. 
In fact, Ge was used to realize the first working transistor and was also among the first materials being intensively studied for surface states and their impact on the electrical properties of a semiconductor \cite{Brattain1953,Hanrath2005,Sistani2020a}. This detailed knowledge is another important benefit of using Ge as plasmonic-loading material in Ge/metal SLSPP structures. However, Ge forms a native germanium oxide with low electrical and chemical stability \cite{Xie2012,Berghuis2021}, together with a poor interface towards the Germanium underneath. This prevented the widespread use of Ge in integrated (optical) circuits for a long time \cite{Loscutoff2006}. The oxide is dominated by the presence of GeO$_2$ \cite{Onsia2004} which is removable in water. Regrettably, it also etches bulk Germanium through a recursive process: first the dominant GeO$_2$ is removed from the surface, followed by surface re-oxidation and another oxide removal in the water. The intermediate surface oxidation step includes the oxidation of water-stable and therefore previously remaining suboxides of GeO$_x$ with x $<$ 2 to GeO$_2$ \cite{Onsia2004}. We will show, that this mechanism is also removing our Ge-SLSPP waveguides and limits their unprotected application in aqueous liquids. \\
However, there is a possible solution to this problem, coming from transistor devices: for being able to use Ge for improving state-of-the-art MOSFETs, different dielectrics have been extensively studied and implemented to cover or substitute the unsuitable native surface oxide(s) of Ge. While first focused attention was given to HfO$_2$ \cite{Zhu2004,Xie2012} due to its highly suitable characteristics and maturity level \cite{Bohr2007}, other promising dielectrics were investigated as well. They include e.g. Al$_2$O$_3$ \cite{Xie2012,Berghuis2021}, TiO$_2$ \cite{Xie2012} and ZrO$_2$ \cite{Tsipas2008,Xie2012} which were all evaluated in our study. Figure \ref{Waveguide+Sim}(a) shows a sketch of the layer structure of the SLSPP waveguides including surface protection layer ("Passivation layer" in the sketch), inside an external measurement setup. It includes a mid-IR QCL to excite the plasmon in the waveguide and a Mercury-Cadmium-Telluride (MCT) detector for measuring the signal. Figure \ref{Waveguide+Sim}(b) displays the corresponding cross-section profile of the waveguide including typical device dimensions.

\section{Plasmonic Waveguide Fabrication}
\begin{figure}[ht]
						\centering
						\includegraphics[width=1\linewidth]{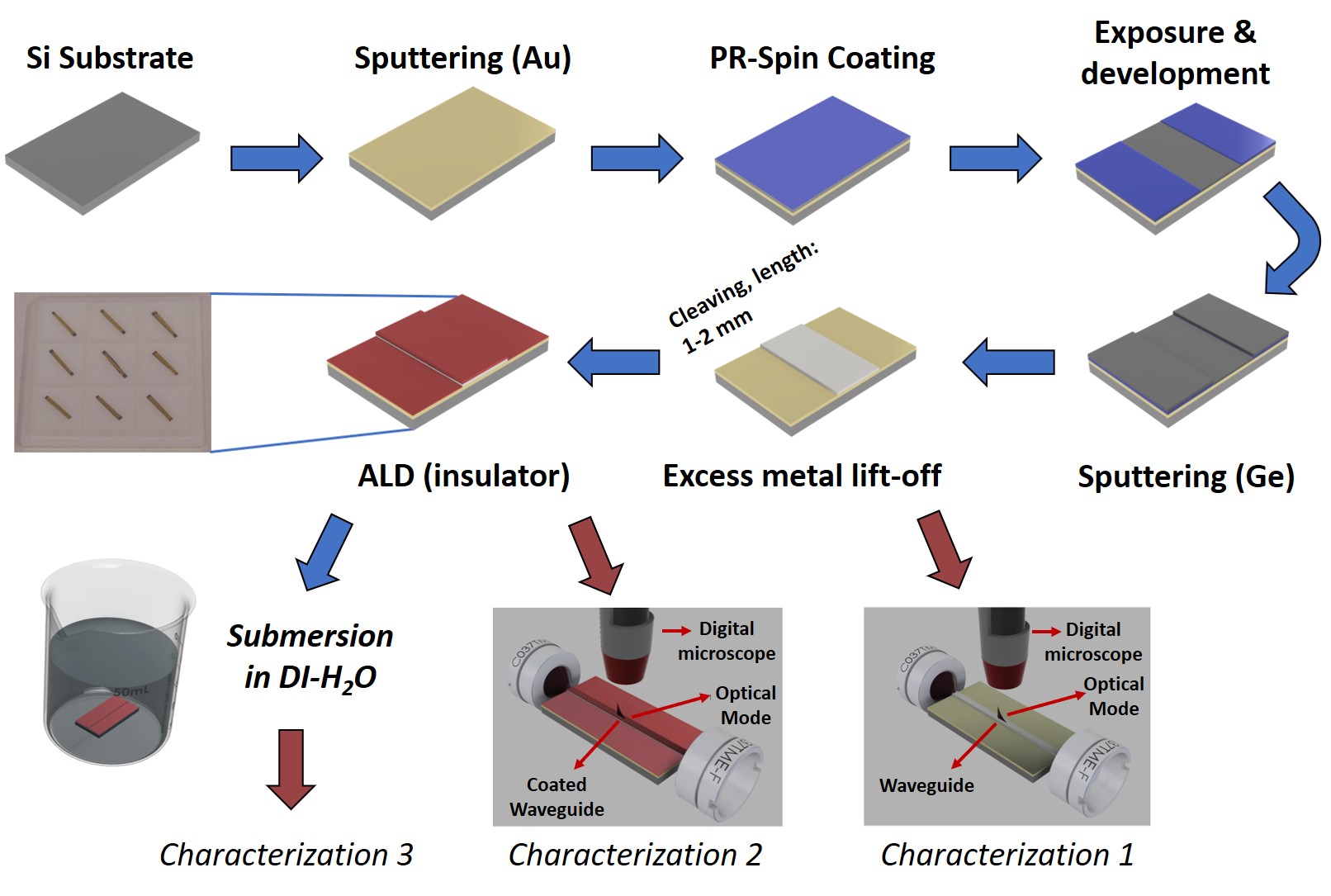}
						\caption{Process flow of the Ge/Au plasmonic waveguide fabrication including the deposition of the ALD-based passivation layers in a post-processing step. The pre-, post-ALD passivation and post-water-submersion characterization (1 - 3) of the samples is indicated. The corresponding custom made waveguide setup is shown in Fig. \ref{exp_setup}.}
						\label{fabrication}
\end{figure}

The Ge/Au SLSPP plasmonic waveguides were fabricated according to the flow chart in Fig. \ref{fabrication} on a double-side polished 275~$\mu$m thick Si (100) substrate. First, a 100~nm thick Au layer is sputtered on the substrate (7 cycles, P$_{base}\leq1e^{-5}$~mbar), together with a 10~nm Ti sticking-layer between metal and semiconductor. After lithography and development, the 300~nm thick Ge surface-loading layer is sputtered on top of the gold (12 cycles, P$_{base}\leq1e^{-5}$~mbar). In contrast to previous work, where the needed Ge-slab was patterned by reactive ion etching (RIE) \cite{David2021}, we defined the 9~$\mu$m wide waveguides in this work through a lift-off process. This results in waveguides with slightly structurally degraded sidewalls and therefore increased line edge roughness, as compared to previous similar structures using a RIE etching process for Ge patterning \cite{David2021} (see e.g. a SEM picture of a typical fabricated Ge ridge in Fig. S2(a) of the supplemental material). Some fabrication defects from the lift-off process can be identified, but they show negligible impact on the losses of the waveguides. The overall ridge smoothness and sidewall quality of the devices are still fairly good, which enables conducting the coating study as intended. This will also be confirmed by the following characterization of the typical waveguide losses, which are on the same level as in our previous work.

\noindent The devices were cleaved to various waveguide lengths between 1~mm and 2~mm and characterized in the previously mentioned waveguide setup (see. Fig. \ref{exp_setup}, \textit{Characterization 1}) to obtain their average waveguide and coupling losses. These reference values are needed for the quantitative analysis of the following steps including the ALD deposition and water submersion experiment. \\
After this initial characterization, individual chips were each coated with nominally 10 nm of one of the four different protective ALD-materials Al$_2$O$_3$ (10.2 nm), TiO$_2$ (9.8 nm), HfO$_2$ (8.5 nm) and ZrO$_2$ (10.1 nm). While aiming for a reasonable layer thickness of about 10 nm to combine proper encapsulation with limited additional losses, especially the HfO$_2$ turned out thinner than expected, i.e. $\sim$8.5~nm. After these coatings were deposited, the waveguides were characterized again (\textit{Characterization 2}) in the waveguide setup. Finally, the submersion experiment was conducted, followed by another sample characterization (\textit{Characterization 3}).

\section{Plasmonic Waveguide Characterization}
\begin{figure}[ht]
						\centering
						\includegraphics[width=1\linewidth]{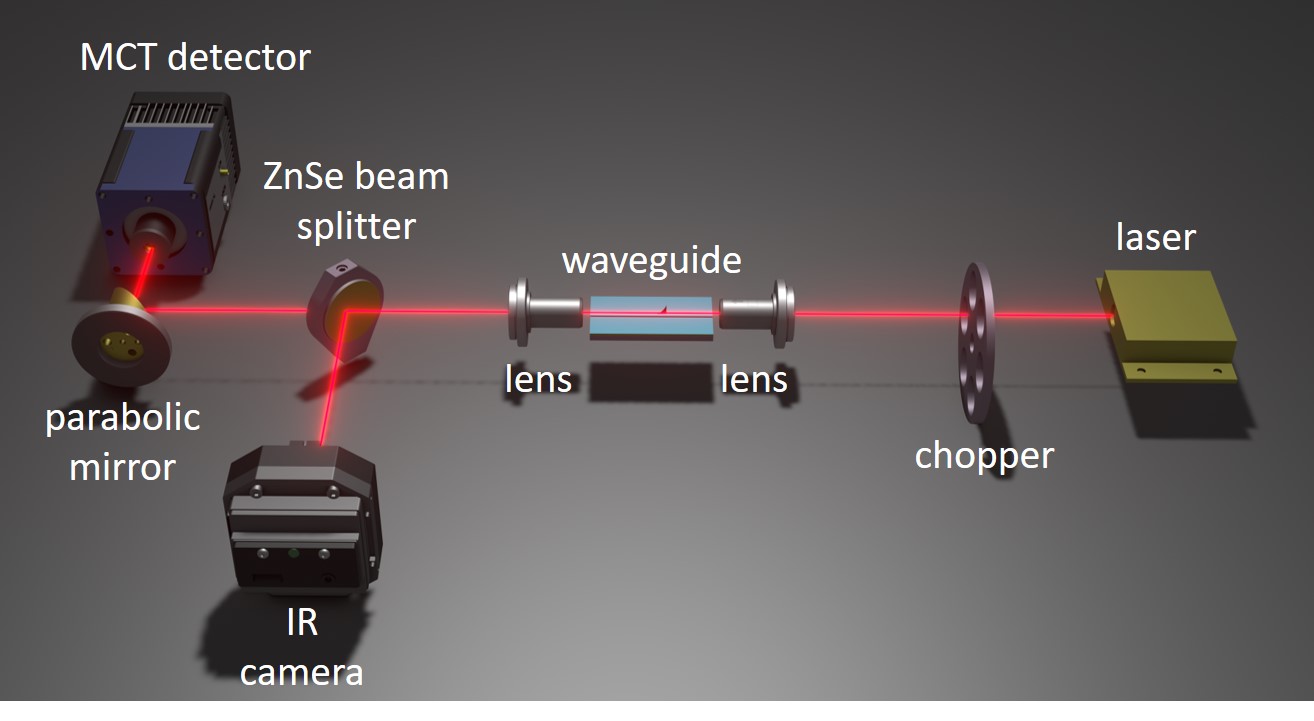}
						\caption{Sketch of the custom made waveguide characterization setup using a 9.38 $\mu$m emitting singlemode DFB QCL. The optically chopped CW laser beam is focused onto the front facet of the Ge/Au-based plasmonic waveguide with a suitable lens (effective focal length: 1.873 mm). After propagating along the waveguide surface it is out-coupled at the back facet and collimated with another identical lens. The waveguide is positioned using piezo-actuators with nanometer precision. The beam is then separated with a ZnSe beam splitter, where one beam is analyzed on a mid-IR camera (\textit{Tau 2}, Teledyne FLIR LLC, USA) for its 2D profile while the other is focused on a thermoelectrically cooled MCT detector. A lock-in amplifier was used as it strongly improved the signal-to-noise-ratio.}
						\label{exp_setup}
\end{figure}

A sketch of the custom-made waveguide characterization setup is shown in Fig. \ref{exp_setup}. We characterized the waveguides using a continuous wave (CW) singlemode distributed feedback (DFB) QCL emitting at a wavelength of 9.38 $\mu$m (mirSense, France). The laser was operated at an optical output power of 1.2 mW in CW mode. The laser beam is collimated and coupled into the plasmonic waveguide using a focusing (in-coupling) and a collimating (out-coupling) lens optics, after passing through a beam chopper. For nanometer precision alignment, the waveguides are positioned on a piezo-actuator stage and follow a dedicated alignment routine, as described previously \cite{David2021}. The out-coupled beam from the waveguides is then either directed to a mid-IR camera (\textit{Tau 2}, Teledyne FLIR LLC, USA) for inspecting the beam profile, or directly focused onto a thermoelectrically cooled MCT detector (\textit{PVI-4TE-10.6}, Vigo Systems S.A., Poland) by a parabolic mirror. The measured signal is filtered for optimized signal-to-noise ratio using a lock-in amplifier. In order to obtain the individual waveguide losses pre- and post-ALD-coating as well as after

\noindent the submersion experiment in water, we first characterized the typical coupling losses of our SLSPP waveguides. As described in previous works \cite{Nikolajsen2003,Zektzer2014,David2021}, the effective cut-back method can be used to extract the waveguides losses. For this, we measure the losses of various waveguides as a function of their length. 
\noindent The obtained total waveguide losses (material plus scattering/fabrication losses) amount for 9.66 dB/mm. All these values agree well with our previous findings \cite{David2021}.

\section{Passivation of mid-IR SLSPP Waveguides \label{sect_passivation}}
Working with semiconductor-based devices, especially when using heterostructures or other nanometer thin layer sequences, requires a high degree of control of the interfaces between the different materials. This is especially true, when using surface-sensitive structures such as SLSSP waveguides. Simultaneously protecting their surface while maintaining the (plasmonic) functionality poses one of the key challenges. Thus, the main requirements concerning coatings for mid-IR plasmonics used in liquid sensing applications are: a) a layer that prevents "physical" penetration from the surrounding liquid, b) low additional optical losses and c) a preservation of the plasmonic characteristics. All three criteria can be satisfied very well by using high-quality dielectric materials known from CMOS-transistors. They are on the nanometer-scale and: a) homogeneously encapsulate the Ge-slab below, b) show (material-dependent) relatively low losses, and c) do not interfere significantly with the underlying plasmonic structure as they are only nanometer-thick dielectrics. The materials we tested are the well-established and previously mentioned Ge-compatible dielectrics HfO$_2$, ZrO$_2$, Al$_2$O$_3$ and TiO$_2$.
\begin{figure}[ht]
						\centering
						\includegraphics[width=0.75\linewidth]{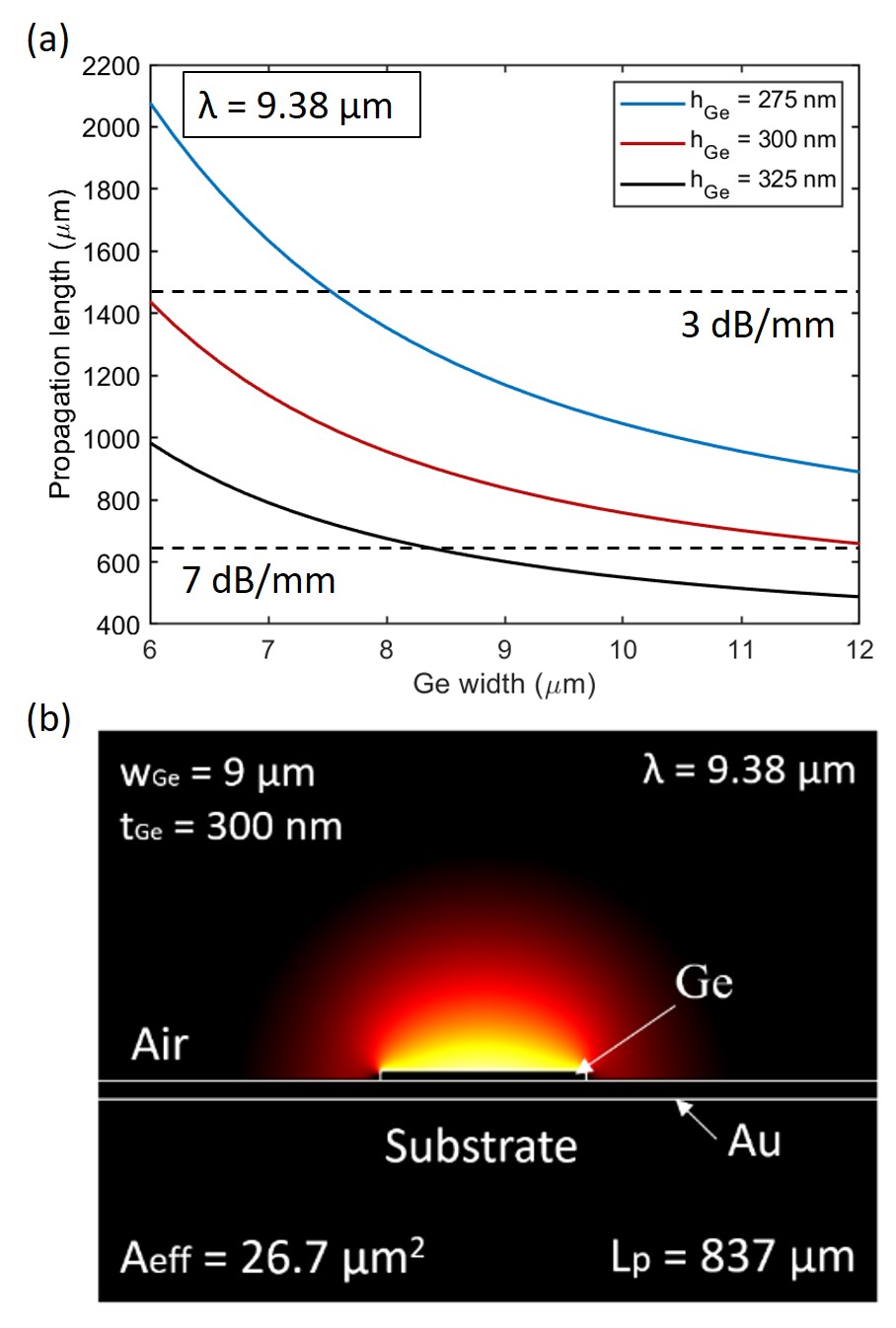}
						\caption{(a) Simulated propagation length as function of the Ge-slab width of the plasmonic waveguide for three different slab thicknesses at $\lambda$ = 9.38 $\mu$m. The crossing point of 3 dB/mm losses (50\% losses) and 7 dB/mm losses (80\% losses) are indicated for orientation. (b) FEM-based simulation of the mode cross-section for a typical Ge/metal based waveguide at $\lambda$ = 9.38 $\mu$m. Calculated propagation length L$_\textrm{P}$ 	
										 and effective mode area A$_\textrm{eff}$ are given.}
						\label{prop-length}
\end{figure}

\noindent Before conducting the experiments, we evaluated the additional waveguide losses of the nanometer-scale passivation layers by FEM-simulations. As a starting point, Fig. \ref{prop-length} shows the simulations of the bare Ge SLSPPs at 9.38~$\mu$m without protective layer. They rely on ellipsometer measurements of the complex mid-IR refractive index of Ge and Au and are given in the supplemental material A, Fig. S1(a) and (b). \\
In Fig. \ref{prop-length}(a) we see the modal propagation length as a function of the Ge-slab width for 3 different Ge-thicknesses. As expected and previously observed \cite{David2021}, an increasing Ge width results in an increased lateral mode confinement and, therefore, higher modal losses from the Au layer.  It is worth noticing, that the propagation length changes significantly with the Ge slab thickness. If thinner Ge stripes provide longer propagation, the consequent increase in mode size can introduce coupling losses with other coupled optical elements (in this case, lenses) \cite{David2021}. It follows, that for sensing platforms it is strongly beneficial to keep thickness variations to a minimum, to not increase the modal losses. On the other hand, increasing the Ge-thickness from 275 nm (blue) to 325 nm (black) also reduces the propagation length for the same confinement reason. For sensing, 9 $\mu$m wide and 300 nm thick Ge-slabs are a good compromise between low losses (reasonable propagation length) and good modal overlap with the surrounding probe medium. The corresponding mode cross-section for this geometry is shown in Fig. \ref{prop-length}(b), demonstrating the significant overlap of $>$ 95\% of the mode with the surrounding medium. The propagation length is still almost a millimeter in this case. \\
For simulating the impact of the additional protective coatings and due to the lack of literature on the complex mid-IR refractive index of the coating materials \cite{Kischkat2012}, we first performed mid-IR ellipsometry measurements on unstructured 

\begin{figure}[ht]
						\centering
						\includegraphics[width=0.75\linewidth]{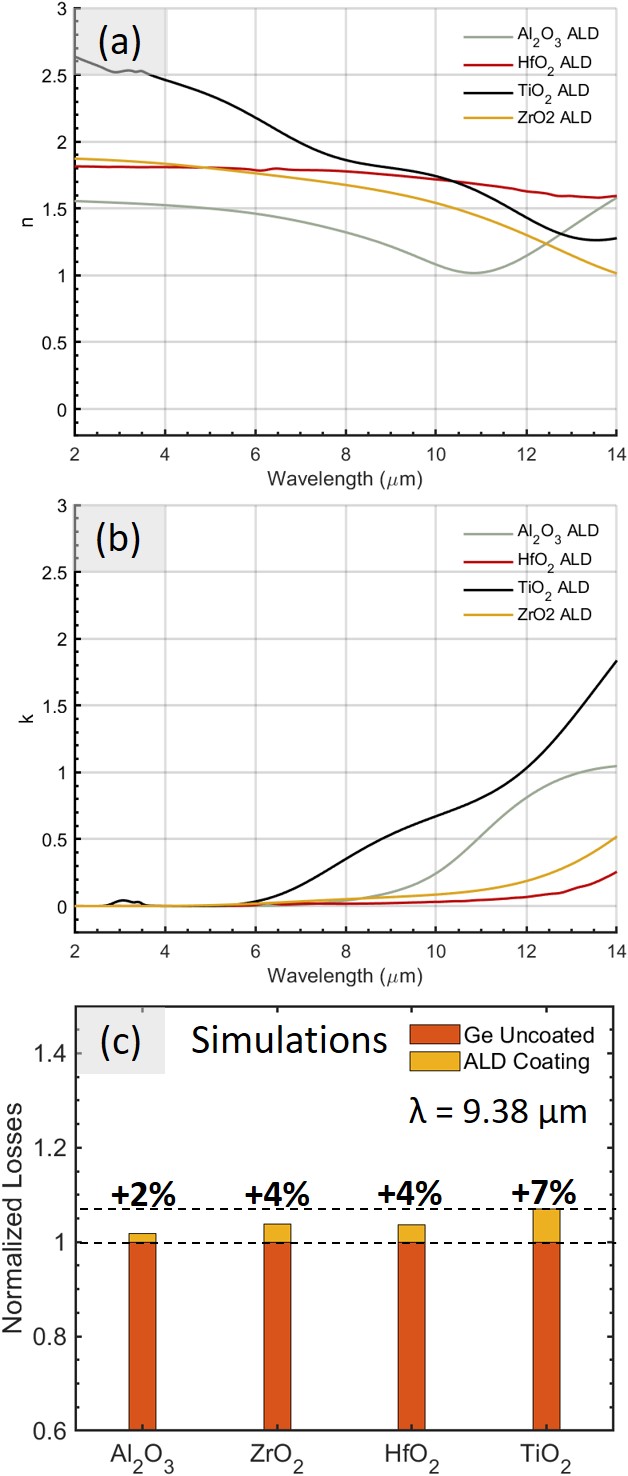}
						\caption{Spectrally dependent mid-IR (a) real part and (b) imaginary part of the refractive index of the four different ALD coatings, i.e. Al$_2$O$_3$ (green), HfO$_2$ (red), TiO$_2$ (black) and ZrO$_2$ (yellow), obtained from mid-IR ellipsometry measurements. (c) Corresponding additional losses to the waveguides, when simulating the impact of applying 10 nm thick layers of the different coating at 9.38 $\mu$m wavelength.}
						\label{coating-sim}
\end{figure}

\noindent Au/Ge/Ge-oxide stacks. The results are shown in Fig. \ref{coating-sim}(a) and (b). Figure \ref{coating-sim}(c) shows the resulting additional losses at 9.38~$\mu$m, simulated including the four different passivation layers. They are displayed as relative increase compared to simulating the bare Ge-slabs on Au without those protective coatings. The different passivation layers add between 2\% (Al$_2$O$_3$) and 7\% (TiO$_2$) to the losses of a typical Ge/Au SLSPP waveguide, compared to without them. These supplemental losses can be divided into additional absorption losses from the respective imaginary part of the refractive index of the coating shown in Fig. \ref{coating-sim}(b) and into the mode confinement-dependent Au-losses, which can be derived from the real part of the refractive index n of the coating displayed in Fig. \ref{coating-sim}(a). As explained above, a higher real(n) leads to a stronger mode confinement and, thus, increased Au-losses. Even though the additional losses are not fully negligible, the coatings are still well suitable for plasmonic applications in the mid-IR. Especially, when considering state-of-the-art plasmonic liquid sensors with their very short needed plasmonic sensing sections on the order of $\sim$10 - 50~$\mu$m \cite{Fabian2006,Schwaighofer2021,Hinkov2022}, our presented propagation lengths on the order of $\gg$ 500 $\mu$m are highly suitable for such applications. \\
High-$k$ materials are typically grown or deposited on the substrate by chemical vapor deposition (CVD) or ALD. The latter intrinsically combines monolayer-thickness control with excellent conformality of the deposited films \cite{Ritala2002} as well as a high layer uniformity \cite{Bethge2009}, making it the deposition technique of our choice. The four protective coatings HfO$_2$, ZrO$_2$, Al$_2$O$_3$ and TiO$_2$ in our study were deposited using a commercial ALD reactor (\textit{Savannah}, Cambridge NanoTech Inc, USA). The samples were inserted into the reactor and thermalized for 10 minutes before starting the process. A silicon substrate was added alongside the waveguide samples and used as a reference for the precise measurement of the deposited layer thickness. A detailed description of the ALD process is given in the supplemental material section B. \\
An initial surface characterization of the fabricated (un)coated SLSPPs yields smooth Ge-surfaces with moderate sidewall roughness. Indeed, the roughness shows a certain variation along the waveguides and a first analysis indicates that it is somewhat more pronounced for the Al$_2$O$_3$ and ZrO$_2$ coated waveguides. More details are given in the supplemental material section C and in Fig. S2 therein.

\section{Submersion Experiment in H$_2$O}
For analyzing the degree of protection provided by the additional ALD coatings, we performed a submersion-experiment in pure DI-H$_2$O. Since the Ge-slabs of the waveguides are relatively thick ($\sim$300 nm) the dissolution process is expected to take place on the minutes to hour(s) timescale \cite{Harvey1958,Henderson2008}. In total we tested 6 samples: 2 uncoated reference samples with bare Ge waveguides only, and the 4 additional ALD-coated waveguide samples. Details on all samples can be found in Tab. I, including the root-mean-square (RMS) roughness from post-submersion AFM measurements.

\begin{table}[ht]
\begin{center}
\caption{Investigated devices including two uncoated Ge references and the four ALD coated samples ZrO$_2$, Al$_2$O$_3$, TiO$_2$ and HfO$_2$. They are compared with respect to their ALD deposition thickness and their RMS roughness values after submersion into H$_2$O for 90 minutes.}
\begin{tabular}{| c | c | c |}
\hline
	& \multicolumn{2}{c |}{\bfseries{uncoated}} \\
\hline
	&  \bfseries{Ge 1} & \bfseries{Ge 2} \\
\hline
	\bfseries{ALD thickness} $(nm)$ & --  & -- \\
\hline
	\bfseries{RMS roughness} $(nm)$ & 1.08 & -- \\
\hline
\end{tabular}

\begin{tabular}{c}

\end{tabular}

\begin{tabular}{| c | c | c | c | c |}
\hline
	& \multicolumn{4}{c|}{\bfseries{ALD coating material}} \\
\hline
	&  \bfseries{ZrO$_2$} & \bfseries{Al$_2$O$_3$} & \bfseries{TiO$_2$} & \bfseries{HfO$_2$} \\
\hline
	\bfseries{ALD thickness} $(nm)$ & 10.1 & 10.2 & 9.8 & 8.5 \\
\hline
	\bfseries{RMS roughness} $(nm)$ & 1.01 & 2.53 & 1.56 & 1.10   \\
\hline
\end{tabular}
\end{center}
\label{Tab_ALD}
\end{table} 

\newpage

\begin{figure}[ht]
						\centering
						\includegraphics[width=1\linewidth]{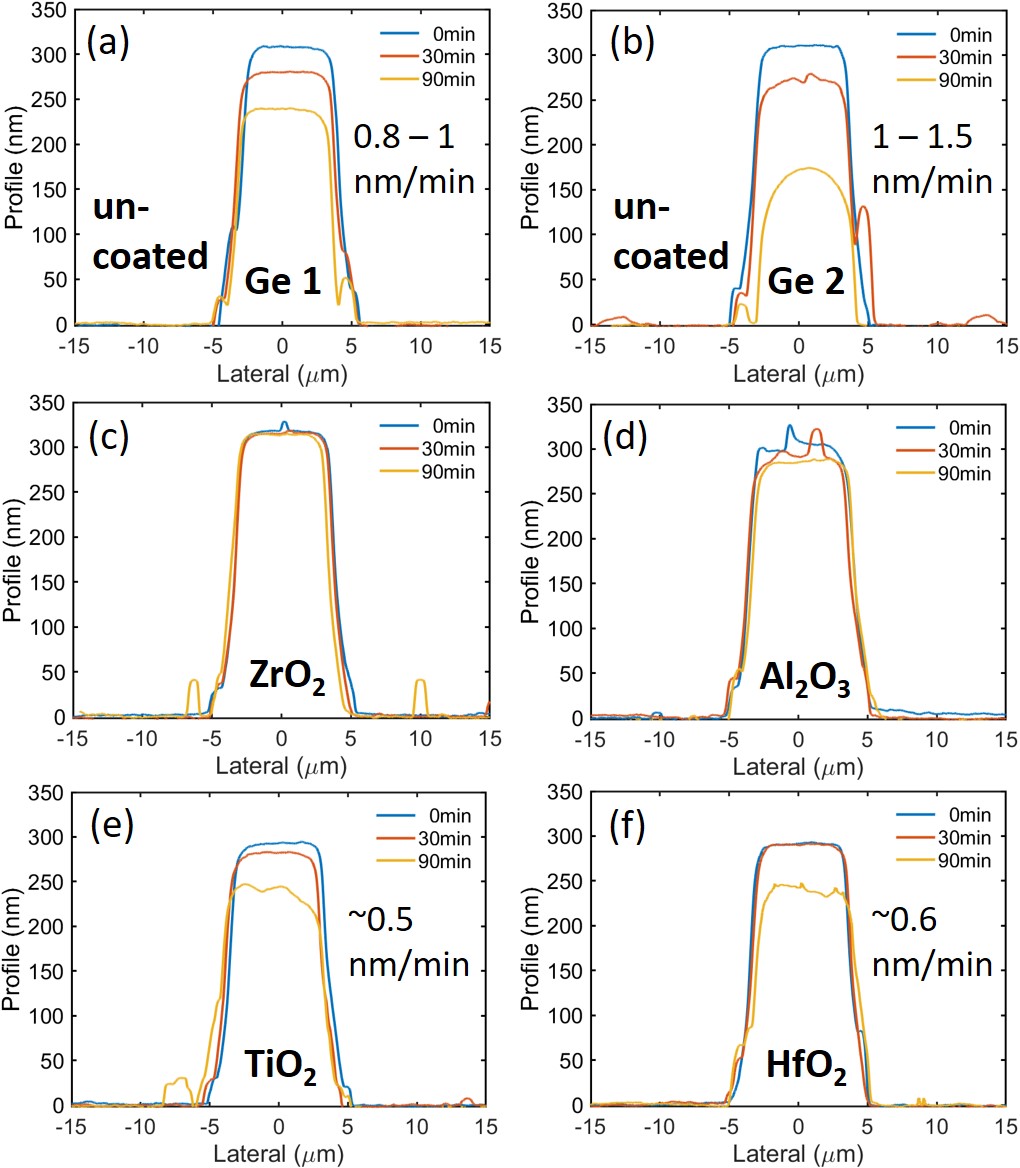}
						\caption{Profilometer measurements for the water submersion experiments (pre: blue, +30 minutes: red, +60 minutes: yellow) of: (a)+(b) uncoated and (c) ZrO$_2$ (d) Al$_2$O$_3$ (e) TiO$_2$ and (f) HfO$_2$ coated waveguides. The uncoated Ge gets clearly dissolved in water, while TiO$_2$ and HfO$_2$ only encounter less pronounced surface degradation. In contrast, the ZrO$_2$ and Al$_2$O$_3$ coated SLSPPs do not show any substantial degradation effects from the water exposure.}
						\label{exp_dektak}
\end{figure}

\noindent The submersion experiment was performed in 2 steps. First the samples were submerged into DI-H$_2$O for 30 minutes and then characterized by optical microscope and by surface profilometry. This was followed by a second submersion for additional 60 minutes (i.e. in total 90 minutes) and the waveguides were analyzed again. Figure \ref{exp_dektak} compares the surface profiles before (blue), after 30 minutes (red) and after 90 minutes (yellow) of submersion. All samples have an initial thickness of about 300 nm. But already after 30 minutes of water exposure, the uncoated samples show a significant reduction in the Ge layer thickness by about 30 nm, i.e. $\sim$10\% of the initial thickness, while in contrast, all coated samples maintain a constant thickness. After additional 60~minutes, the uncoated samples display another, much stronger, removal of the Ge and sample \# 2 (Fig. \ref{exp_dektak}(b)) is even reduced to about half of its initial thickness ($\sim$150 nm). In addition, the profile of the waveguide changes from rectangular waveguides to a round-shaped geometry. The different etch-rates for the two uncoated devices originate most likely from non-identical surface conditions before water exposure, being a result from variations in the fabrication process. \\
After 90 minutes of submersion, the ZrO$_2$ and Al$_2$O$_3$ coatings remain unaffected, while TiO$_2$ and HfO$_2$ show first degradation effects by a reduced layer thickness of about 50 nm. For HfO$_2$ this can be explained by a thinner deposited layer thickness of $\sim$8.5 nm, while for TiO$_2$ AFM measurements reveal holes in the coating (see Fig. S4 in the supplemental material). Still the coatings give a certain protection, demonstrated by no thickness reduction within the first 30 min and by the effective etch rate that can be calculated after 90 min from Fig. \ref{exp_dektak}: while for the uncoated Ge-slabs the etch-rate is between 1 - 1.5 nm$\cdot$min$^{-1}$, it is 0.5 nm$\cdot$min$^{-1}$ and 0.6~nm$\cdot$min$^{-1}$ for TiO$_2$ and HfO$_2$, respectively. \\
An additional issue arises from the lack of a stable Ge surface-oxide, which might explain the etching of the samples with the TiO$_2$ and especially the HfO$_2$ coating. It is the diffusion of Ge, e.g. known in the case of HfO$_2$, into the dielectric coating layer \cite{Wu2004}, resulting in charge trapping and fixed charges \cite{Gusev2004,Frank2006}. Such diffusion processes can oxidize and, thus, contribute to slowly removing the HfO$_2$ passivation itself in water. They can be strongly suppressed by implementing a SiH$_4$ layer between the Ge and HfO$_2$ \cite{Wu2004}. But also surface nitridation and S-passivation can help to improve the Ge surface quality \cite{Xie2012} as well as simple storage of the sample for some time in air. This could be shown by an improved passivation-quality from storing an Al$_2$O$_3$ coated sample for 3 months in air (surface recombination velocity S$_\textrm{eff,max}$ reduces by factor 3-4) \cite{Berghuis2021}. \\
A detailed top-view microscope analysis of all samples reveals that the submersion shows no visible effects on the waveguide surface or width (see Fig. S2 in the supplemental material section C). This is in good agreement with the profilometer measurements, which also show no reduction in the waveguide width during the submersion experiment, even for the removed Ge surfaces.  \\
For advanced surface characterization, we also performed atomic force microscope (AFM) measurements. Scanning an area of 5 x 5 $\mu$m$^2$ on top of the waveguide after the 90 minutes of water submersion yields the RMS roughness values given in Tab. \ref{Tab_ALD}. While they are on the order of $\sim$1-1.5 nm, with the exception of the Al$_2$O$_3$ coating (RMS ~ 2.5 nm), they show no change after submersion (compare Fig. S4(a), no submersion, and (f), after 90 min in water, in the supplemental material section C). It is important to note that there is no correlation between the measured surface roughness after water exposure and observed degradation effects from being in contact with water. More details on the AFM scans are described in the supplemental material section C.

\begin{figure}[ht]
						\centering
						\includegraphics[width=0.75\linewidth]{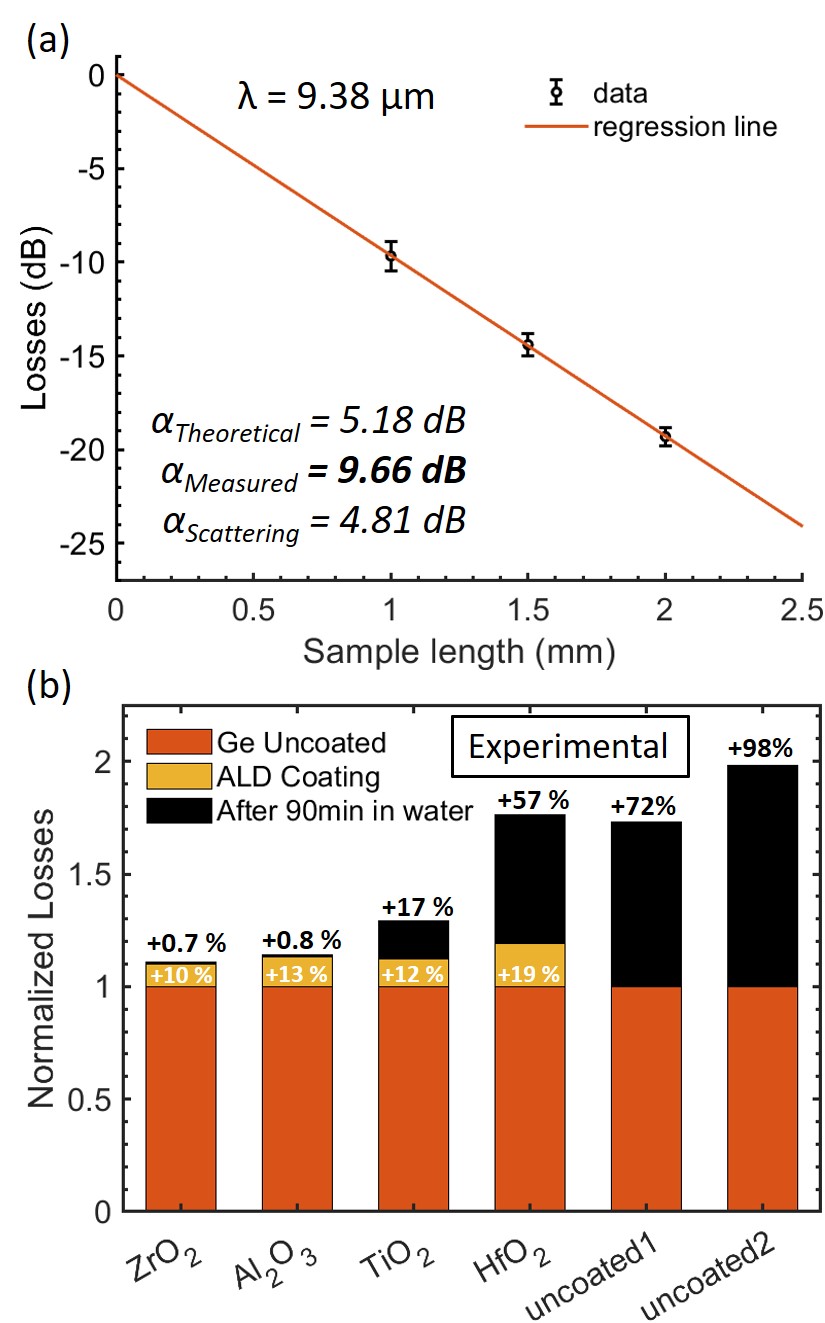}
						\caption{(a) Results of the effective cut-back technique measurements. (b) Normalized experimental waveguide losses before (red), after ALD coatings (yellow) and after the water submersion experiment (black). Depositing the coatings adds roughly 10 - 20\% (on the dB scale) of losses, while the water submersion significantly increases the losses of the uncoated waveguides. But also the losses of the HfO$_2$ coated waveguides and to a minor extend also of the TiO$_2$ coated ones are increased due to too thin or holey coatings for those materials that led to water penetration.}
						\label{exp_losses}
\end{figure}

\noindent Finally, and most importantly, we measured the actual effect of the water exposure on the plasmonic properties of the waveguides at a wavelength of 9.38 $\mu$m. Fig. \ref{exp_losses}(a) displays the total insertion losses in dB as a function of the sample length. This enables extracting the coupling losses as y-intercept and, therefore, to calculate the actual waveguide losses. These measurements were performed before the submersion experiment. We extract scattering/fabrication losses of about 4.81~dB. The final result of the protection study is displayed in Fig. \ref{exp_losses}(b). It shows the evolution of the normalized experimental losses (for better comparison normalized for each device to the losses of its uncoated waveguides) when undergoing the coating process (yellow) and the submersion experiment (black), measured with the setup shown in Fig. \ref{exp_setup}. It is clearly visible that we measure an approximately doubled increase in losses when adding the coatings (from +10\% for ZrO$_2$ to +19\% for HfO$_2$) as compared to the expected simulation results displayed in Fig. \ref{coating-sim}. After 90 min of submersion, we observe that the uncoated samples experience the expected increase in their losses (+72\% and +98\%) due to the significant decrease in layer thickness. In agreement with the previously measured profile change, this effect is even more pronounced for the sample "uncoated2". For the ALD-coated samples we demonstrate excellent protective capabilities for ZrO$_2$ and Al$_2$O$_3$ with no distinct additional waveguide losses ($<$1\%), while TiO$_2$ (+17\%) and even more pronounced HfO$_2$ (+57\%) show increasing losses. This agrees very well with the previous findings in the reduction of total waveguide thickness after water exposure. \\
We can conclude that our approach of covering a Ge-layer by a 10~nm ALD-coating results in a very good protection from Ge-removal over time, when selecting the proper materials (ZrO$_2$ and Al$_2$O$_3$) and thicknesses (10 nm coating thickness and above). We believe that by increasing the thickness of the two so far less successful coatings TiO$_2$ and HfO$_2$, we will be able to achieve a similar level of protection. This can become relevant for certain spectroscopic experiments, where exposing the Ge-based SLSPP waveguides to the analyte might specifically benefit from using TiO$_2$ as a coating \cite{Gong2001}. Its additional properties as a robust and bio-compatible coating material \cite{Hofmann2003} and seed for activated surfaces \cite{Frank2021}, can in such a case counter-balance the expected higher overall losses from a thicker coating.

\section{Surface Activation of ZrO$_2$ Coated SLSPP Waveguides}
\begin{figure}[ht]
						\centering
						\includegraphics[width=1\linewidth]{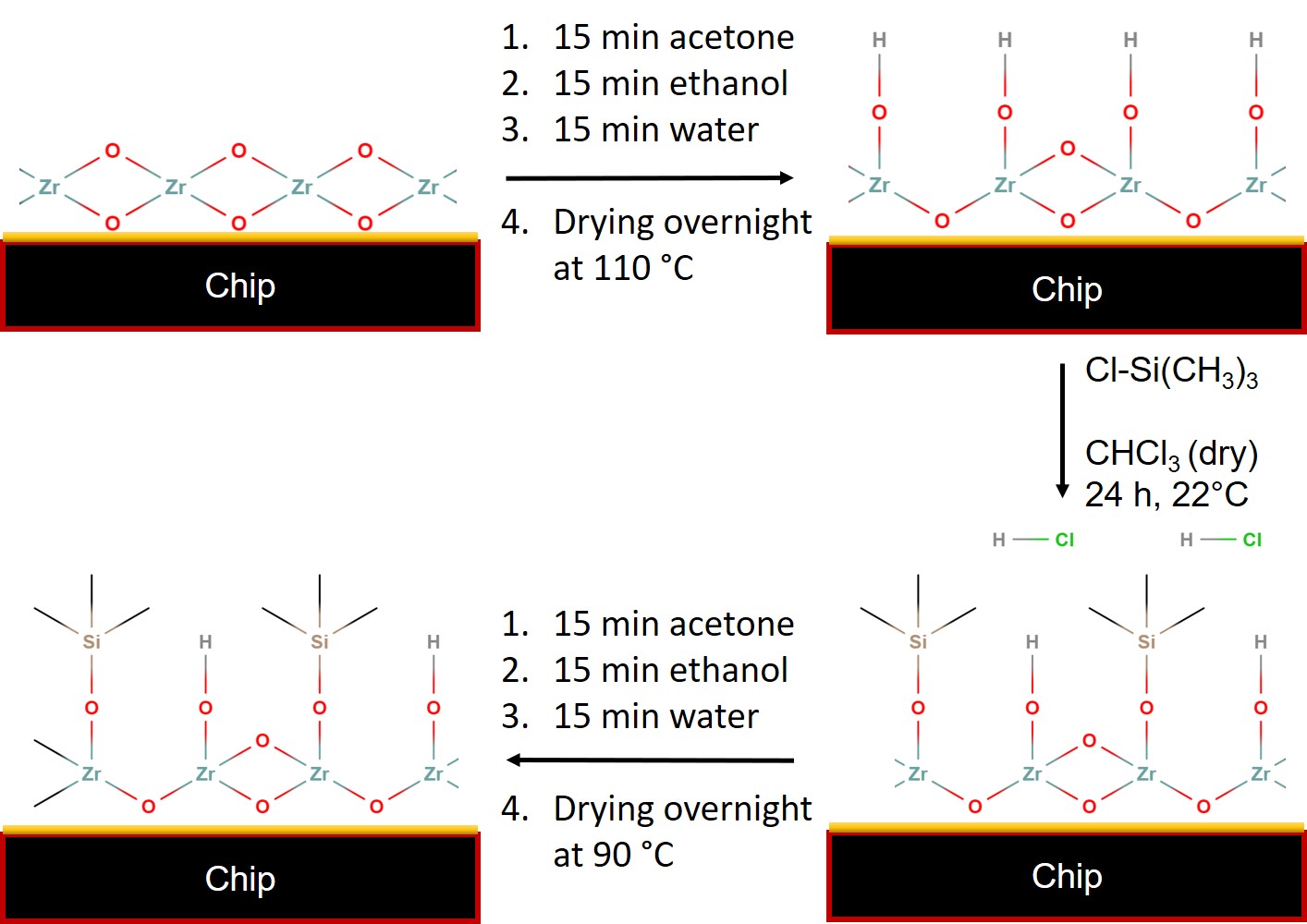}
						\caption{Procedure of the chemical treatment to activate the chip surface of ZrO$_2$ coated plasmonic surfaces. Details are given in the main text.}
						\label{functionalization}.
\end{figure}

Besides simple protective capabilities from additional surface coatings for SPP waveguides, the activation and functionalization of such surface-sensitive interfaces is a powerful tool to realize next generation liquid sensors, where chip-scale footprint as well as selective and sensitive detection features are needed \cite{Mizaikoff2013}. The application of surface-modified waveguides in mid-IR sensing has already been shown in literature, where different coating materials were used ranging from mesoporous materials to metal organic frameworks. The surface modification is used to increase the sensitivity by enriching the analyte in the surface layer, while repelling unwanted substances like water and keeping them away from the evanescent field \cite{Han1998,Wijaya2011,Kim2018,Beneitez2020,Husseini2021}. In addition, these coatings can also be applied in catalysis \cite{Taguchi2005,Bavykina2020}. \\
In this respect, Zirconia, very similar to Titania and Alumina, is a very interesting material, because of its bio-compatibility \cite{Hofmann2003,Affatato2004,Shankar2021,Tchinda2022} and ability to serve as a host layer for surface functionalization strategies, e.g. based on mesoporous coatings \cite{Wacht2021}. Such membranes have been shown to improve the sensitivity of state-of-the-art ATR-sensors by factors of more than 160, which would effectively push recently demonstrated monolithic SPP-based liquid sensors \cite{Schwarz2014,Hinkov2022} from their current limit of detection on the ppm- to the ppb-scale. In addition, ZrO$_2$ shows a high chemical and mechanical stability, which is a favorable property when applying such coatings in sensing schemes \cite{Nawrocki1993}. \\
In contrast to merely using the implemented passivation coatings as a simple protection layer in our waveguides, we want to show another possible approach for activating the plasmonic surface. The idea is to keep unwanted water away from the evanescent field of the light traveling through the plasmonic waveguide, while simultaneously increasing the selectivity of our surface layer towards certain molecules. As the pristine ZrO$_2$ surface is usually hydrophilic, we added a trimethylsilyl functionality to the surface to introduce a higher affinity towards hydrophobic compounds, similar to the shown mesoporous ZrO$_2$ layers on Si-ATR crystals \cite{Wacht2021}. In a first test, the surface of one of the SLSPP waveguide samples, which was covered with 12 nm of ZrO$_2$, was functionalized as described in literature \cite{Wacht2021}. Unfortunately, this waveguide experienced damages to its surface and the procedure had to be re-done without the use of the ultrasonic bath. Instead, elongated periods of submerging the waveguides in the solvents were used, revealing no damages to the processed surface by microscope inspection. Consequently, the procedure displayed in Fig. \ref{functionalization} was used to functionalize the surface of the SLSPP waveguides, using the reagents and materials described in D of the supplemental material: \\
The sample was submerged in acetone, ethanol and deionized water for 15 min each. After purging the waveguides with dry air, the sample was put into a drying oven at 110$^{\circ}$C overnight. Then, the sample was placed in a three-neck round bottom flask with a reflux condenser and bubbler, a N$_2$-inlet and a vacuum connection. It was dried for 2 h at approximately 15 mbar and 150~$^{\circ}$C. After that, the flask was purged with N$_2$ and 20 mL of CHCl$_3$ and 400 $\mu$L of chlorotrimethylsilane at 22$^{\circ}$C were added. This mixture was kept in an inert atmosphere for 24 h. Then, the solution was removed, the waveguide sample thoroughly washed with CHCl$_3$ and subsequently submerged in acetone, ethanol and deionized water for 15 min each. It was finally purged with dry air and then placed overnight in a drying oven at 90$^{\circ}$C. The success of the surface modification was determined by comparing the contact angles of a water drop on waveguides with and without surface functionalization.

\noindent Details on the contact angle measurement routine are given in section D.2 of the supplemental material. As seen in Fig. \ref{functionalization1}(a), the non-functionalized ZrO$_2$-surface shows a lower contact angle compared to the functionalized ZrO$_2$-surface. A lower contact angle corresponds to a better wetting of the surface and thus higher hydrophilicity. After the surface functionalization procedure, the contact angle is significantly increased, which corresponds to a higher hydrophobicity. This change is a first indication of a successful introduction of the trimethylsilyl moiety to the ZrO$_2$ surface. In addition, the losses of the plasmonic mode of the waveguide caused by the functionalization of the surface were compared to a non-functionalized waveguide. Figure \ref{functionalization1}(b) clearly shows the higher absorption losses caused by the trimethylsilyl-functionalization with long-term exposure to elevated temperatures, which is expected to lead to gold diffusion into its surrounding layers together with a change in crystallinity of the Ge for the same temperature reason. In addition, the formed Zr-O-Si bonds after functionalization could also increase the measured losses. This is another strong indication for a successful modification of the ZrO$_2$-surface. \\
Finally, the adsorption capabilities of an organic, apolar analyte on the pristine and functionalized ZrO$_2$-surface were investigated. Diethyl ether was used as such an analyte, since its FTIR spectrum shows absorption bands at the wavelength of interest of 9.38 $\mu$m. Moreover, it is a volatile compound that evaporates rapidly unless dissolved or adsorbed to a surface. To investigate the adsorption capabilities of the waveguides, we used the same waveguide setup as described in Fig. \ref{exp_setup}. First, the plasmonic losses of both the non-functionalized and functionalized waveguide were determined experimentally. Then, each waveguide was submerged in diethyl ether for 10 min and the plasmonic losses were measured again. The difference between the waveguide losses before and after diethyl ether submersion define the relative losses caused by the absorption of the adsorbed apolar analyte. The such determined plasmonic losses are displayed in Fig. \ref{functionalization1}(b). When comparing them for 

\begin{figure}[ht]
						\centering
						\includegraphics[width=0.7\linewidth]{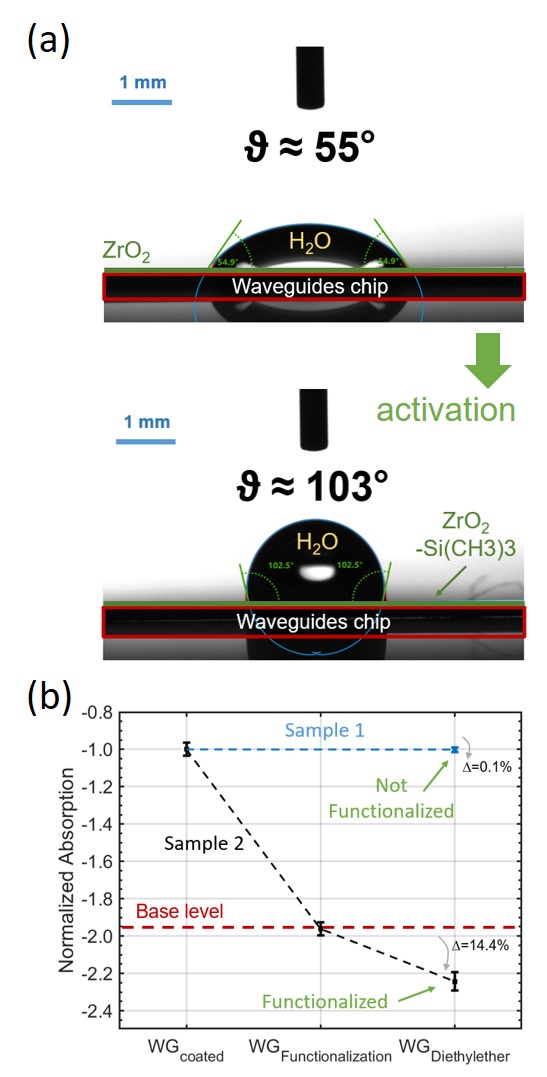}
						\caption{(a) Contact angle measurements, using a water droplet on the non-activated (top, 3 $\mu$L, hydrophilic: $\vartheta$ $<$ 90$^{\circ}$) and activated (bottom, 7 $\mu$L, hydrophobic: $\vartheta$ $>$ 90$^{\circ}$) SLSPP waveguide surface. (b) Results of the surface activation measurements on the activated ZrO$_2$ plasmonic surfaces for: no functionalization (Sample 1) and functionalization (Sample 2). While submerging the not functionalized waveguides into diethyl ether results in no additional losses for the plasmonic mode, the functionalized surface shows increased losses of about 14.4 \% because of the molecules sticking to the surface which absorb the 9.38 $\mu$m probe light.}
						\label{functionalization1}.
\end{figure}

\noindent the non-functionalized waveguide (Sample 1) before and after diethyl ether submersion, almost no difference is visible. This means, that the analyte evaporates before it can be measured with the setup, indicating no adsorption to the surface.
When instead doing the same experiment with the functionalized waveguides (Sample 2), a distinct difference is visible. It indicates surface adsorption of the diethyl ether, which results in higher plasmonic losses from the additional absorption of radiation by the diethyl ether. The relative increase of the plasmonic losses amount for 14.4\%. \\
Interestingly, after approximately 5 min, the analyte completely evaporates again from the surface, resulting in the same plasmonic losses as before submerging the waveguides into the diethyl ether. This indicates a complete recovery of the surface, which is a necessary feature for real-life applications of sensors. \\
The performed experiment clearly demonstrates the suitability of applying surface activated ZrO$_2$ coated SLSPP waveguides for spectroscopic analysis. Together with the wide variability of different silanes available for surface modification and functionalization, the selectivity of these waveguides can be tailored to satisfy different application scenarios. In addition, a higher sensitivity can be obtained by increasing the film thickness or introducing porosity \cite{Wacht2021}. In combination with the compact design of the waveguides and their availability for monolithic integration on the chip-scale \cite{Schwarz2014,Hinkov2022}, it makes them highly suitable for the next generation of liquid sensors with miniaturized footprint and significantly enhanced selectivity and sensitivity.

\section{Conclusion}
In conclusion, we investigated the suitability of broadband Ge-based mid-IR SLSPP waveguides for applications in liquid spectroscopy. While Ge gets dissolved in water over time, we show that protective high-quality ALD coatings can prevent this effect. In particular, approximately 10~nm thick ZrO$_2$ and Al$_2$O$_3$ coatings on the Ge-slabs withstand an exposure to water for 90 minutes and maintain the plasmonic capabilities of the waveguide, as demonstrated by following optical waveguide measurements. The successful additional coatings increase the absolute losses by about 10\% - 13\%, i.e. 1-2 dB$\cdot$mm$^{-1}$. In contrast, TiO$_2$ and HfO$_2$ show worse protection, and the Ge waveguides get significantly removed after 90 minutes in water. This can be explained by holes in the TiO$_2$ coating as seen in the AFM measurements and a too thin protective layer for the HfO$_2$, together with Ge/Ge-oxide diffusion into the protective layer, making it susceptible to oxidation and its removal in water. \\
We further show at the example of the ZrO$_2$ coatings, that the surface can be chemically functionalized, while maintaining its plasmonic properties. In a proof-of-concept experiment after surface functionalization, we were able to reproducibly detect diethyl ether at the SLSPP waveguide surface at a wavelength of 9.38 $\mu$m, identified from additional measured losses. Additionally, we observe an evaporation of the diethyl ether after 5 min, resulting in a similar surface as before the submersion into the diethyl ether. \\

\section*{Acknowledgments}
\noindent Fruitful discussions with W. Schrenk are gratefully acknowledged. The authors want to thank A. Linzer, S. Dal Cin, F. Pilat, A. Dabrowska and G. Maisons for technical support. \\
The authors acknowledge funding from the EU Horizon 2020 Framework Program under grant agreement numbers 780240 and 828893. B.H. received funding from the Austrian Science Fund FWF (M2485-N34). We want to thank the CzechNanoLab Research Infrastructure supported by MEYS CR (LM2018110) for the expert technical assistance.  


\bibliographystyle{IEEEtran}
\bibliography{David_et_al._ALD_Plasmonic_Coatings}

\end{document}